\documentclass [14pt]{article}
\usepackage{geometry} 
\usepackage[cp1251]{inputenc}
\usepackage[english,russian,ukrainian]{babel}
\usepackage{graphicx}

\begin{document}

\textbf{Deuteron: analytical forms of wave function and density
distribution}

V.I. Zhaba

Uzhhorod National University, 88000, Uzhhorod, Voloshin Str., 54

\textbf{Introduction}: The deuteron wave function can be presented as the
table through the respective massifs of numerically values of radial wave
functions. In calculations such arrays of numbers in the texts of programs
to operate quite difficult. Therefore a used simpler analytical forms is
expedient.

\textbf{Purpose}$: $In the received analytical forms of deuteron wave function
in coordinate representation it is necessary to calculate density
distribution in the deuteron and transition density and compare them among
themselves.

\textbf{Results}$: $On the received coefficients of the four analytical forms
for deuteron wave function for the nucleon-nucleon potential Reid93 are
calculated density distribution and transition density. For comparison,
similar results are given for Moscow potential.

\textbf{Conclusion}$: $Such calculations help to evaluate the correctness of the
choice of the analytical form for approximation of the wave function, and
also to obtain information about such characteristics of the deuteron as
charge form factor, tensor polarization and momentum distribution.

\textbf{Key words:} deuteron, wave function, analytical form, approximation,
density distribution, transition density.

\textbf{Дейтрон: аналитические формы волновой функции и
распределение плотности}

В.И. Жаба

Ужгородский национальный университет, 88000, Ужгород, ул. Волошина, 54

По полученным ранее коэффициентам четырех аналитических форм
волновой функции дейтрона в координатном представлении для
нуклон-нуклонного потенциала Reid93 рассчитано распределение
плотности в дейтроне и плотность перехода. Такие расчеты помогают
оценить корректность выбора аналитической формы для аппроксимации
волновой функции, а также получить информацию о таких
характеристиках дейтрона как зарядовый форм-фактор, тензорная
поляризацию и распределение импульсов.

\textbf{Ключевые слова:} дейтрон, волновая функция, аналитическая форма,
аппроксимация, распределение плотности, плотность перехода.

УДК 539.12.01

PACS 03.65.Nk, 13.88.+e, 21.45.Bc

\textbf{Дейтрон: аналітичні форми хвильової функції та розподіл
густини}

В.І. Жаба

Ужгородський національний університет, 88000, Ужгород, вул.~Волошина, 54

По отриманим раніше коефіцієнтам чотирьох аналітичних форм
хвильової функції дейтрона в координатному представленні для
нуклон-нуклонного потенціалу Reid93 розраховано розподіл густини в
дейтроні та густину переходу. Такі розрахунки допомагають оцінити
коректність вибору аналітичної форми для апроксимації хвильової
функції, а також одержати інформацію про такі характеристики
дейтрона як зарядовий формфактор, тензорна поляризацію та розподіл
імпульсів.

\textbf{Ключові слова:} дейтрон, хвильова функція, аналітична форма,
апроксимація, розподіл густини, густина переходу.

\textbf{Вступ}

Дейтрон є найпростішім ядром. Він складається з двох елементарних
частинок - протона і нейтрона. Простота і наочність будови
дейтрона робить його зручною лабораторією для вивчення і
моделювання нуклон-нуклонних сил. На сьогодні дейтрон добре
вивчений експериментально і теоретично. Експериментально визначені
значення статичних характеристик дейтрона добре узгоджуються з
експериментальними даними. Однак незважаючи на це, існують певні
теоретичні неузгодженості і проблеми. Наприклад, в деяких роботах
хвильова функція дейтрона (ХФД) в координатному представленні має
вузли поблизу початку координат \cite{1}. Існування таких вузлів
основного і єдиного стану дейтрона свідчить про неузгодженості і
неточності в реалізації чисельних алгоритмів в розв'язанні
подібних задач або про особливості потенціальних моделей дейтрона.

Також слід відмітити, що такі потенціали NN взаємодії, як
Боннський, Парижський, Московський, потенціали Неймегенської групи
\cite{2}, Argonne v18, NLO, NNLO та N$^{3}$LO, Idaho N$^{3}$LO чи
Oxford мають досить непросту структуру і доволі громіздкий запис.
Оригінальний потенціал Reid68 був параметризований Неймегенською
групою на основі фазового аналізу і отримав назву Reid93.

Крім того, ХФД може бути представлена таблично: через відповідні масиви
значень радіальних хвильових функцій. Іноді при чисельних розрахунках
оперувати такими масивами чисел доволі складно і взагалі незручно. І текст
програм для чисельних розрахунків є громіздкий, перевантажений і
нечитабельним. Тому є доцільним отримання більш простих аналітичних форм
представлення ХФД. У подальшому по них можна розрахувати формфактори і
тензорну поляризацію, що характеризують структуру дейтрона.

ХФД у зручній формі необхідні для використання у розрахунках як
поляризаційних характеристик дейтрона, так і для оцінки теоретичних значень
спінових спостережуваних в \textit{dp}- розсіянні.

У даній роботі розглядається вплив вибору аналітичних форм при апроксимації
на величини розподілу густини та густину переходу в дейтроні.

\textbf{Аналітичні форми ХФД}

В 2000-x рр. були запропоновані нові аналітичні ХФД в
координатному представленні. Серед них слід відмітити
параметризації Дубовиченко і Бережного-Корда-Гах \cite{3}, а також
параметризацію у виді \cite{4}

\begin{equation}
\label{eq1}
\left\{ {\begin{array}{l}
 u(r) = r\sum\limits_{i = 1}^N {A_i e^{ - a_i r^2},} \\
 w(r) = r\sum\limits_{i = 1}^N {B_i e^{ - b_i r^2};} \\
 \end{array}} \right.
\end{equation}

У роботі \cite{5} наявна ще така апроксимація із застосуванням
функцій Лагерра $\psi _{3n} $

\begin{equation}
\label{eq2}
\left\{ {\begin{array}{l}
 u(r) = \sum\limits_{n = 0}^N {A_n \psi _{3n} (r),} \\
 w(r) = \sum\limits_{n = 0}^N {B_n \psi _{3n} (r),} \\
 \end{array}} \right.
\end{equation}

Крім форм (\ref{eq1}) і (\ref{eq2}), у роботах \cite{6} та
\cite{7} використано такі аналітичні форми відповідно

\begin{equation}
\label{eq3}
\left\{ {\begin{array}{l}
 u(r) = r^{3 / 2}\sum\limits_{i = 1}^N {A_i e^{ - a_i r^3},} \\
 w(r) = r\sum\limits_{i = 1}^N {B_i e^{ - b_i r^3}.} \\
 \end{array}} \right.
\end{equation}

\begin{equation}
\label{eq4}
\left\{ {\begin{array}{l}
 u(r) = r\sum\limits_{i = 1}^N {A_i e^{ - a_i r},} \\
 w(r) = r^3\sum\limits_{i = 1}^N {B_i e^{ - b_i r}.} \\
 \end{array}} \right.
\end{equation}

Причому вибір останньої зумовлений асимптотиками ХФД: $\chi _l (r) \sim r^{l
+ 1}$на малих відстанях та $\chi _l (r) \sim e^{ - \alpha r}$при $r \to
\infty $.

Незважаючи на громіздкі і довготривалі розрахунки і мінімізації \textit{$\chi $}$^{2}$ (до
величини менших за 10$^{ - 4})$, за допомогою формул (\ref{eq1})-(\ref{eq4}) проводилась
апроксимація чисельних значень ХФД для потенціалів Неймегенської групи (в
тому числлі і для Reid93). Масиви чисел становили 839х2 значень для
радіальних ХФД в інтервалі $r$=0-25~fm.

Коефіцієнти аналітичних форм (\ref{eq1})-(\ref{eq4}) для
потенціалу Reid93 приведені в \cite{4, 6, 7, 8} відповідно.
Розраховані ХФД не містять надлишкових вузлів поблизу початку
координат. Отримані по ХФД статичні параметри і поляризаційні
характеристики добре узгоджуються з літературними теоретичними й
експериментальними даними.

\textbf{Розподіл густини в дейтроні}

ХФД є двокомпонентною \cite{9}:

\begin{equation}
\label{eq5}
\Psi _d^{M_d } (r) = R_0 (r)Y_{011}^{M_d } + R_2 (r)Y_{211}^{M_d } ,
\end{equation}

де $R_{0}=u/r$; $R_{2}=w/r$ -- радіальні функції S- та D- станів; $Y_{LSJ}^{M_d } $
- спін-кутові функції.

У короткому діапазоні структура дейтрона наочно описується
розподілом густини \cite{9} (або нуклонним розподілом густини
речовини в дейтроні \cite{3}) $\rho _d^{M_d } (r',\theta )$, який
залежить від проекції $M_{d}$ повного кутового моменту, відстані
$r'$ від центра мас і полярного кута $\theta $ до $r'$. Стандартне
нормування залежить від міжчастинкової відстані $r = 2r'$

\[
\int\limits_0^\infty {r^2\left[ {R_0^2 (r) + R_2^2 (r)} \right]dr} = 1;
\]

\[
\int {\rho _d^{M_d } (r')d^3r'} = 2.
\]

Використовуючи (\ref{eq5}), для проекцій $M_{d}$=0;$\pm $1
отримують розподіл густини $\rho _d^{M_d } $ у виді \cite{9}

\begin{equation}
\label{eq6}
\left\{ {\begin{array}{l}
 \rho _d^0 = \frac{4}{\pi }\left[ {C_0 (2r') - 2C_2 (2r')P_2 (\cos \theta )}
\right]; \\
 \rho _d^{\pm 1} = \frac{4}{\pi }\left[ {C_0 (2r') + C_2 (2r')P_2 (\cos
\theta )} \right]; \\
 \end{array}} \right.
\end{equation}

де $C_0 = R_0^2 + R_2^2 $; $C_2 = \sqrt 2 R_0 R_2 -
\frac{1}{2}R_2^2 $ -- компоненти розподілу густини; $P_{2}$ --
поліном Лежандра. При кутах $\theta _1 = 0$ і $\theta _2 = \pi /
2$ очевидно, що справедлива тотожність $\rho _d^0 (r',\theta _2 =
\pi / 2) = \rho _d^{\pm 1} (r',\theta _1 = 0)$. Крім розподілу
густини $\rho _d^{M_d } $, внутрішня структура дейтрона описується
також і густиною переходу $\rho _{tr}^{\pm 1} $ \cite{9}

\begin{equation}
\label{eq7}
\begin{array}{l}
 \rho _{tr}^{\pm 1} (r') = \frac{2}{\pi }\left\{ {R_0^2 (2r') -
\frac{1}{2}R_2^2 (2r') - } \right. \\
 \left. { - \frac{1}{2}\left[ {\sqrt 2 R_0 (2r')R_2 (2r') + R_2^2 (2r')}
\right]P_2 (\cos \theta )} \right\} . \\
 \end{array}
\end{equation}

На Рис.~1-4 приведено величини розподілу густини (з компонентами
$С_{0}$, $С_{2})$ та густини переходу, отримані по ХФД
(\ref{eq1})-(\ref{eq4}) для потенціалу Reid93. У залежності від
вибору апроксимації ХФД розраховані величини $\rho _d^{M_d } $ і
$\rho _{tr}^{\pm 1} $ відрізняються тільки в області 0-0.25~fm.
Фактично це говорить про те, яка із застосованих апроксимацій є
``кращою'' біля початку координат, незважаючи на відсутність
надлишкового вузла ХФД. Для порівняння на Рис.~5 приведені
величини $\rho _d^{M_d } $ і $\rho _{tr}^{\pm 1} $, розраховані по
ХФД для Московського потенціалу \cite{10}, яка містить надлишковий
вузол для S- хвилі (при $r$=0.65~fm). Тому спостерігається зсув
вправо вузлів та максимумів величин $\rho _i $.

Fig.~1. $\rho _i $ для потенціалу Reid93. ХФД (\ref{eq1})

\pdfximage width 140mm {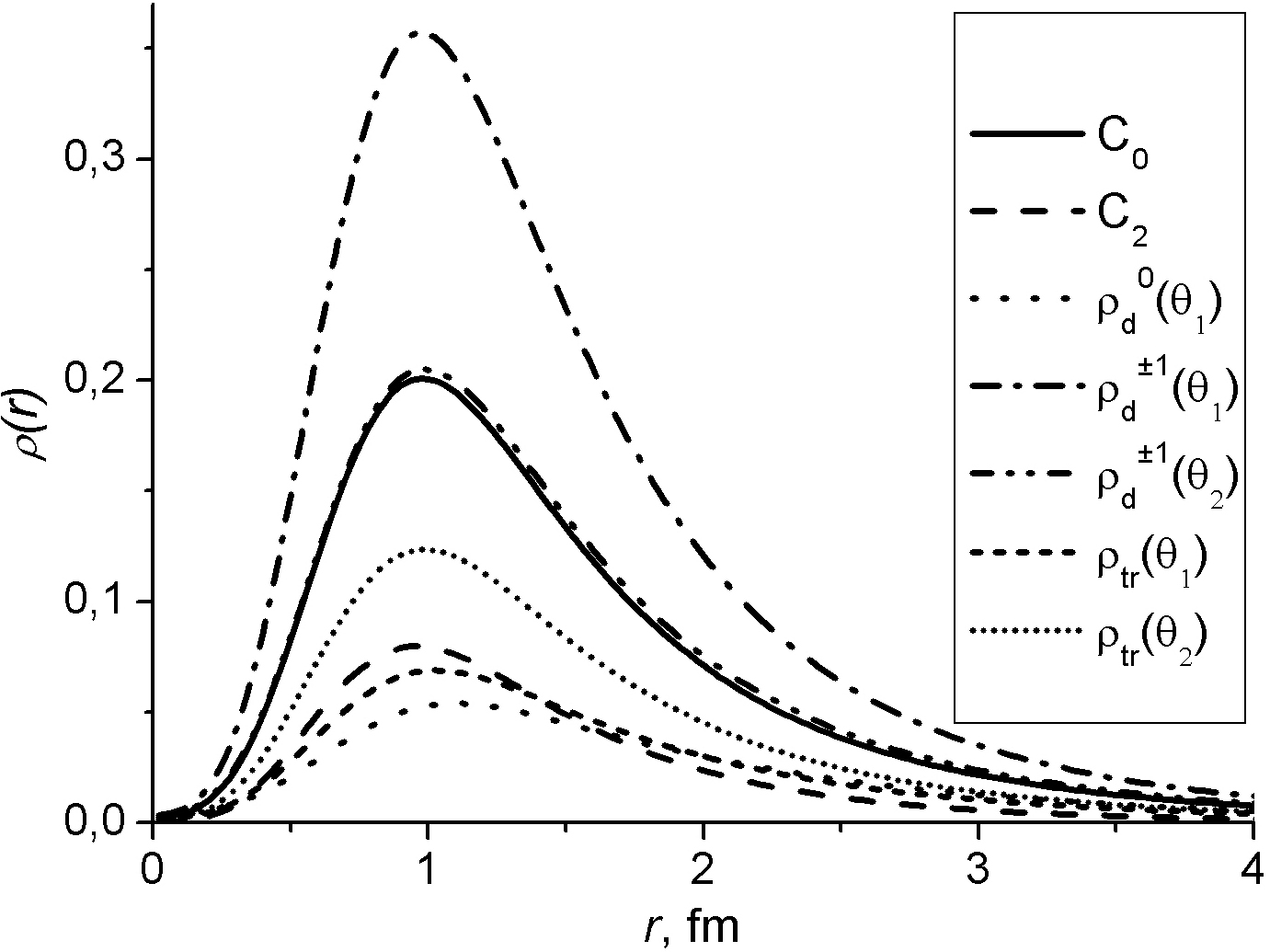}\pdfrefximage\pdflastximage

Fig.~2.  $\rho _i $ для потенціалу Reid93. ХФД (\ref{eq2})

\pdfximage width 140mm {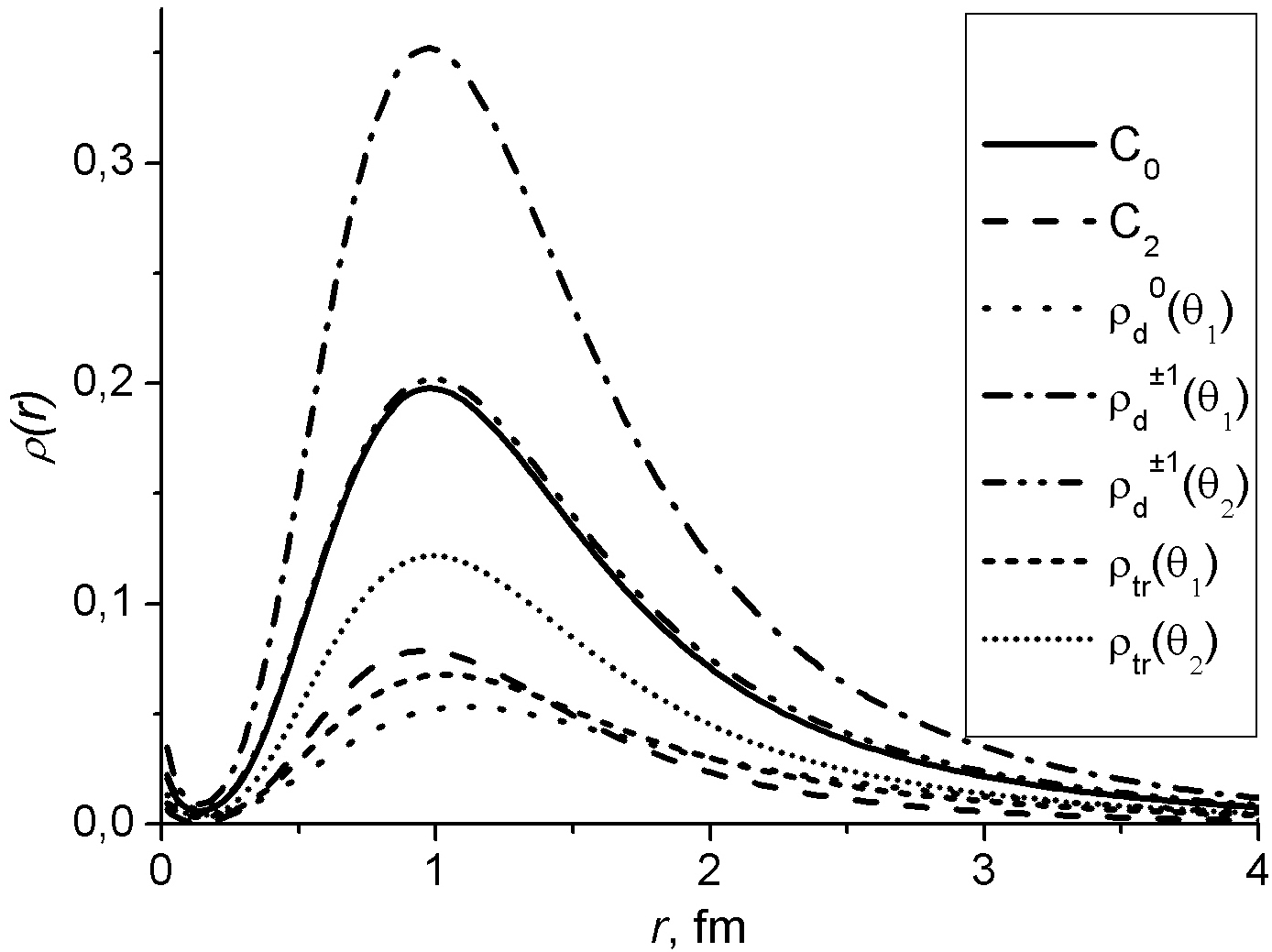}\pdfrefximage\pdflastximage

Fig.~3.  $\rho _i $ для потенціалу Reid93. ХФД (\ref{eq3})

\pdfximage width 140mm {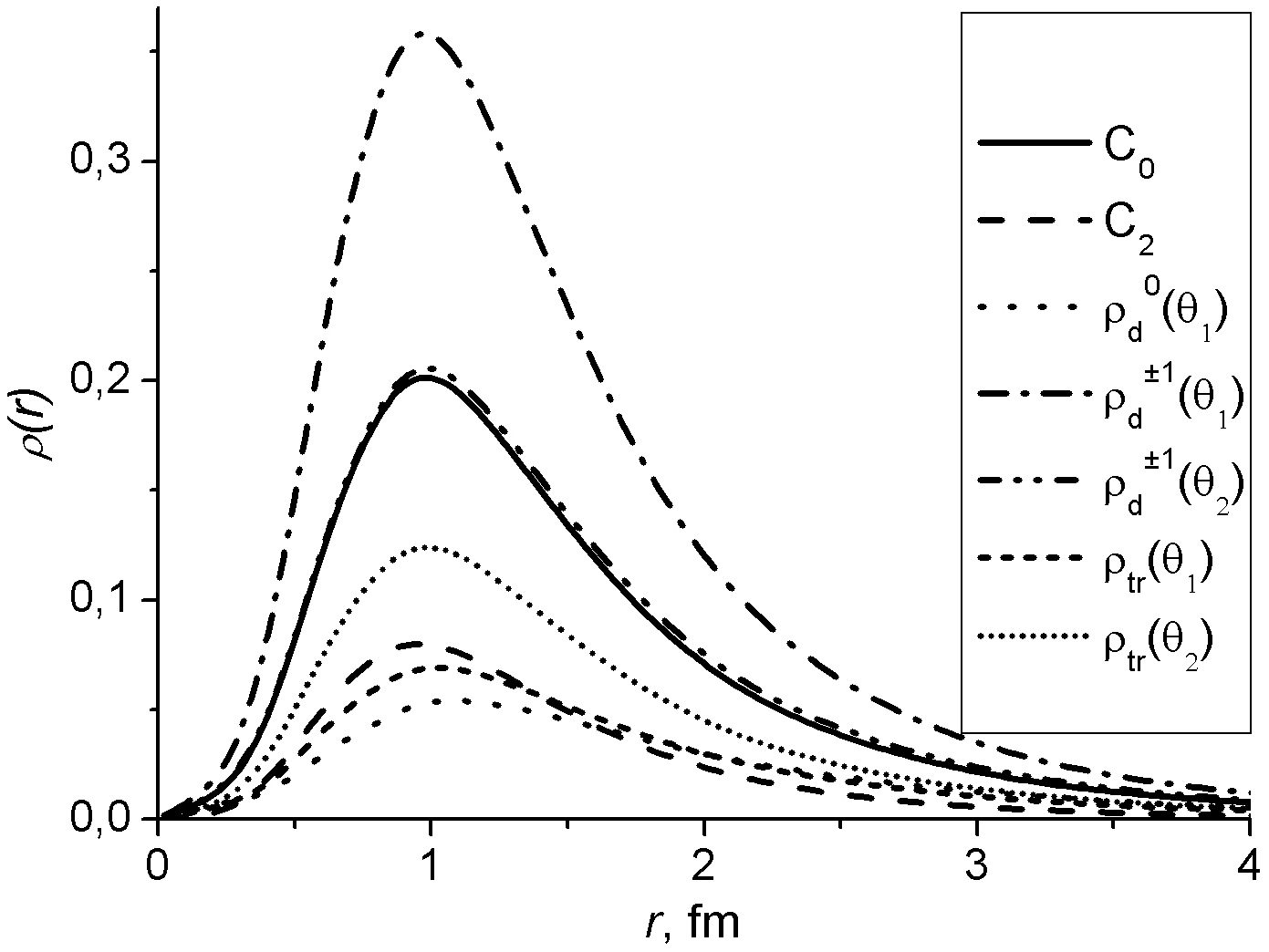}\pdfrefximage\pdflastximage

Fig.~4.  $\rho _i $ для потенціалу Reid93. ХФД (\ref{eq4})

\pdfximage width 140mm {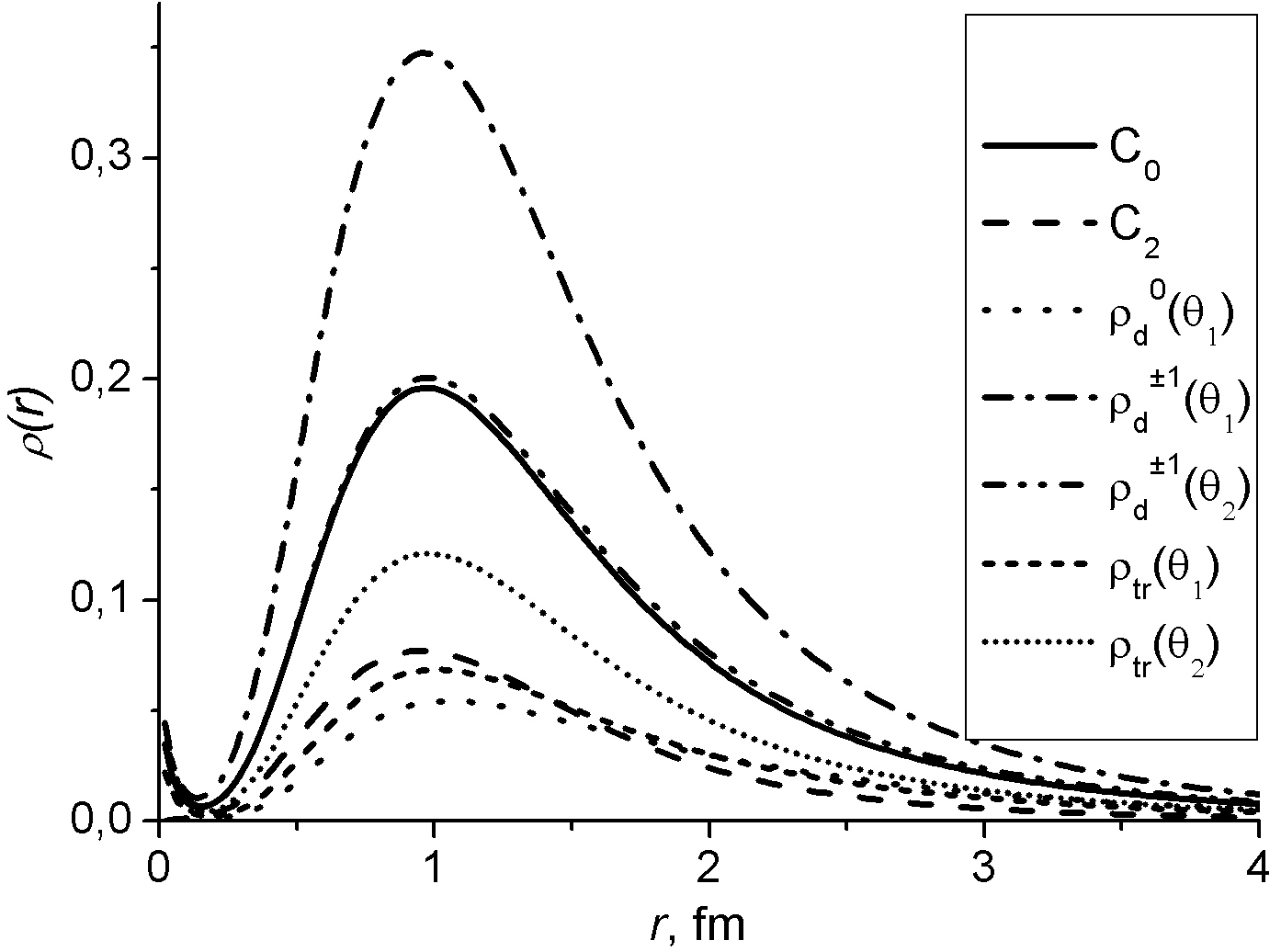}\pdfrefximage\pdflastximage

Fig.~5.  $\rho _i $ для потенціалу для Московського потенціалу

\pdfximage width 140mm {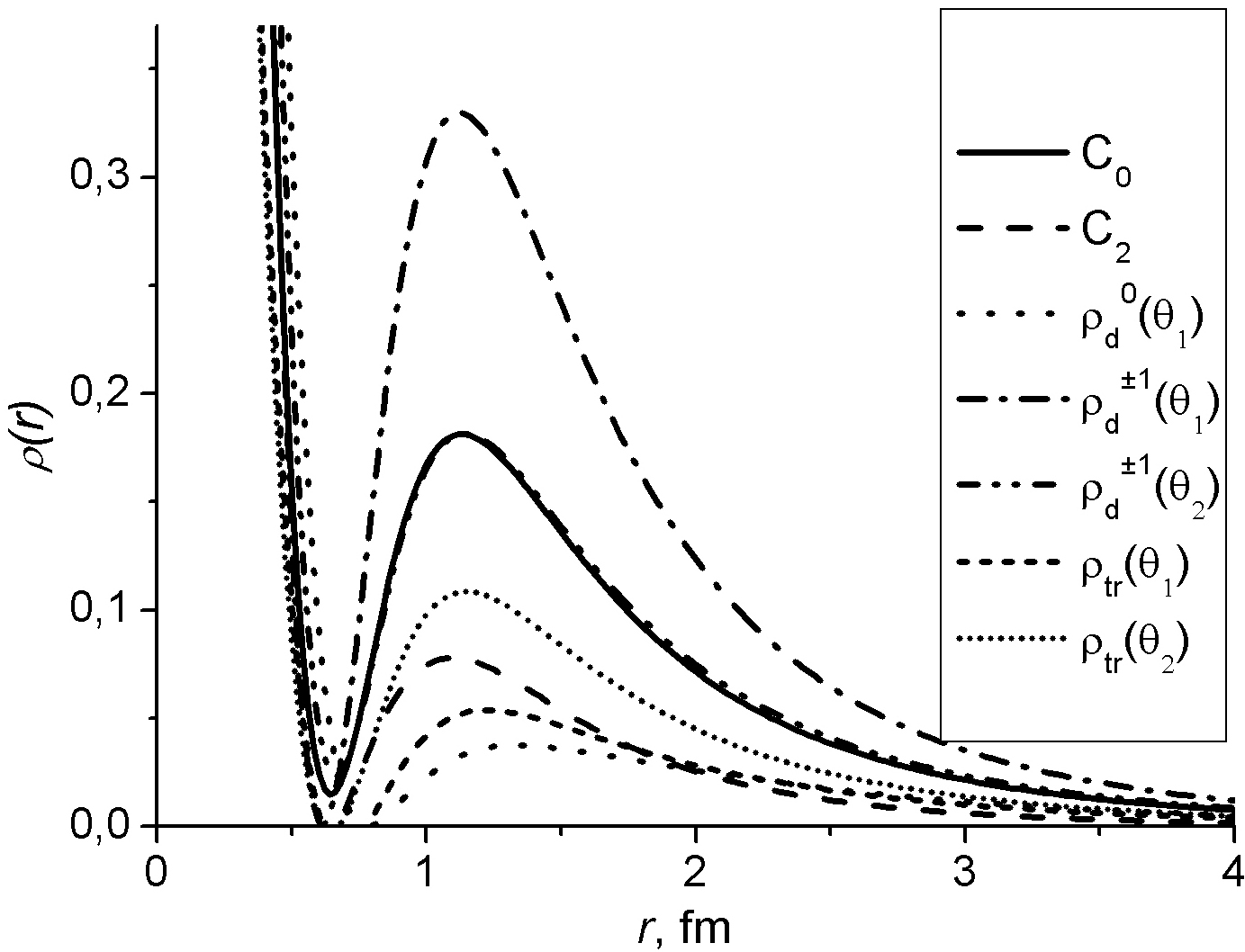}\pdfrefximage\pdflastximage

\textbf{Висновки}

По отриманим раніше коефіцієнтам чотирьох аналітичних форм ХФД (\ref{eq1})-(\ref{eq4}) в
координатному представленні для нуклон-нуклонного потенціалу Reid93
розраховано величини розподілу густини $\rho _d^{M_d } $ та густина переходу
$\rho _{tr}^{\pm 1} $.

Розрахунок величин розподілу густини в дейтроні та густина переходу може
слугувати для оцінки коректності вибору аналітичної форми при апроксимації
ХФД.

Зрештою знання розподілу густини в дейтроні дозволяє отримати
інформацію \cite{9} про його зарядовий формфактор і тензорну
поляризацію. Спін-фліп частина магнітного формфактору та розподіл
імпульсів можуть бути одержані з густини переходу. Це також
забезпечить незалежну оцінку просторових розмірів тороїдальної
структури в дейтроні та перерізу d(e,e'p)n- реакції в одно
фотонному обмінному наближенні.

\end{document}